\documentclass[epj]{webofc}
\usepackage[varg]{txfonts}   
\woctitle{MESON2016 - the 14$^\textrm{th}$ International Workshop on Meson Production, Properties and Interaction}
\begin{document}
\newcommand{\pizerod}{\pi^0_D}
\newcommand{\npcl}{\SI{90}{\percent} CL\xspace}
\selectlanguage{english}
\title{Kaon experiments at CERN: recent results and prospects}

\author{Evgueni Goudzovski\inst{1}\fnsep\thanks{For the NA48/2 and NA62 collaborations:
F.~Ambrosino, A.~Antonelli, G.~Anzivino, R.~Arcidiacono,
W.~Baldini, S.~Balev, J.R.~Batley, M.~Behler, S.~Bifani, C.~Biino
% ...
}
}

\institute{School of Physics and Astronomy, University of Birmingham, B15 2TT, United Kingdom}

\abstract{
The NA48/2 and NA62-$R_K$ experiments at the CERN SPS collected large samples of charged kaon decays in flight in 2003--07. The data analysis is still on-going (with over 20 publications produced so far); the recent results from these experiments are presented. A new upper limit on the rate of a lepton number violating decay
$K^\pm\to\pi^\mp\mu^\pm\mu^\pm$ is reported: $\mathcal{B} < 8.6\times 10^{-11}$ at 90\% CL. Searches for heavy sterile neutrino $N_4$ and neutral scalar resonances ($\chi$) in $K^\pm\to\pi\mu\mu$ decays are reported. Upper limits on the products $\mathcal{B}(K^\pm\to\mu^\pm N_4)\mathcal{B}(N_4\to\pi^\mp\mu^\pm)$ and $\mathcal{B}(K^\pm\to\pi^\pm \chi)\mathcal{B}(\chi\to\mu^+\mu^-)$ are set in the range $10^{-10}$ to $10^{-9}$ for resonance lifetimes up to 100~ps. A preliminary measurement of the electromagnetic transition form factor slope of the $\pi^0$ from $1.05\times 10^6$ fully reconstructed $\pi^0\to\gamma e^+e^-$ decays is presented: the obtained result $a = (3.70 \pm 0.53_\text{stat} \pm 0.36_\text{syst})\times 10^{-2}$ represents the first observation of a non-zero slope in the time-like region of momentum transfer.
}
\maketitle

\newpage

\section*{Introduction}

The NA48/2 experiment at the CERN SPS collected a large sample of charged kaon decays in 2003--04 (corresponding to about $2\times 10^{11}$ $K^\pm$ decays in the vacuum decay volume). The experiment used simultaneous $K^+$ and $K^-$ beams and was optimized for the search for direct CP violating charge asymmetries in the $K^\pm\to3\pi^\pm$ decays~\cite{ba07}. Its successor, the NA62-$R_K$ experiment, collected a 10 times smaller $K^\pm$ decay sample with low intensity beams and minimum bias trigger conditions in 2007 using the same detector~\cite{la13}. The large data samples accumulated by the two experiments have allowed precision studies of rare $K^\pm$ decays and searches for new physics. Recent results from these experiments are reported.

\section{Beam, detector and data sample}
\label{sec:experiment}

The NA48/2 and NA62-$R_K$ experiments used simultaneous $K^+$ and $K^-$ beams produced by 400~GeV/$c$ primary CERN SPS protons impinging on a beryllium target. Charged particles in a narrow momentum band were selected by an achromatic system of four dipole magnets which split the two beams in the vertical plane and recombined them on a common axis. The beams then passed through collimators and a series of quadrupole magnets, and entered a 114~m long cylindrical vacuum tank with a diameter of 1.92 to 2.4~m containing the fiducial decay region.

The vacuum tank was followed by a magnetic spectrometer housed in a vessel filled with helium at nearly atmospheric pressure, separated from the vacuum by a thin ($0.3\%~X_0$) $\rm{Kevlar}\textsuperscript{\textregistered}$ window. An aluminium beam pipe of 158~mm outer diameter traversing the centre of the spectrometer (and all the following detectors) allowed the undecayed beam particles to continue their path in vacuum. The spectrometer consisted of four drift chambers (DCH) with an octagonal transverse width of 2.9~m: DCH1, DCH2 located upstream and DCH3, DCH4 downstream of a dipole magnet that provided a horizontal transverse momentum kick of 120~MeV/$c$ for charged particles (for NA48/2). Each DCH was composed of eight planes of sense wires. The DCH space point resolution was 90~$\mu$m in both horizontal and vertical directions, and the momentum resolution was $\sigma_p/p = (1.02 \oplus 0.044\cdot p)\%$, with $p$ expressed in GeV/$c$ (for NA48/2). The spectrometer was followed by a plastic scintillator hodoscope (HOD) with a transverse size of about 2.4 m, consisting of a plane of vertical and a plane of horizontal strip-shaped counters arranged in four quadrants (each logically divided into four regions). The HOD provided time measurements of charged particles with 150~ps resolution. It was followed by a liquid krypton electromagnetic calorimeter (LKr), an almost homogeneous ionization chamber with an active volume of 7 m$^3$ of liquid krypton, $27~X_0$ deep, segmented transversally into 13248 projective $\sim\!2\!\times\!2$~cm$^2$ cells. The LKr energy resolution was $\sigma_E/E=(3.2/\sqrt{E}\oplus9/E\oplus0.42)\%$, the spatial resolution for an isolated electromagnetic shower was $(4.2/\sqrt{E}\oplus0.6)$~mm in both horizontal and vertical directions, and the time resolution was $2.5~{\rm ns}/\sqrt{E}$, with $E$ expressed in GeV. The calorimeter was followed by a muon system consisting of three scintillator planes, with an iron wall installed in front of each plane. A detailed description of the beamline, detector and trigger is given in Refs.~\cite{fa07,la13,ba07}.

\boldmath
\section{Searches for lepton number violation \& resonances in $K^\pm\to\pi\mu\mu$ decays}
\unboldmath
\label{sec:LFV}

Lepton number violation (LNV) has not been observed. Heavy Majorana neutrinos, which are possible mediators of LNV processes, arise in particular in the $\nu\text{MSM}$ model~\cite{Asaka2005} which introduces three heavy sterile neutrinos. The lightest of them has a mass of $\mathcal{O}({\rm keV})$ and is a dark matter candidate, while the other two have masses ranging from 100~MeV to a few GeV. The model explains baryon asymmetry through CP-violating sterile neutrino oscillations and mixing with active neutrinos, and the low neutrino masses through see-saw mechanism. A simple extension of this model with a scalar field (the inflaton, $\chi$) also explains the homogeneity and isotropy of the universe on large scales and the existence of structures on smaller scales~\cite{Asaka2006}.

A particular case of the LNV decay $K^\pm\to\pi^\mp\mu^\pm\mu^\pm$ has been investigated with the NA48/2 data sample. This decay can be mediated by the sterile neutrinos produced via $K^\pm\to\mu^\pm N$ and subsequently decaying via $N\to\pi^\mp\mu^\pm$. For an off-shell neutrino $N$ with a mass $m_4$, the accessible mass range is $m_\pi+m_\mu < m_4 <m_K-m_\mu$, and the branching fraction for this channel would be
\begin{equation*}
\mathcal{B}(K^\pm\to\pi^\mp\mu^\pm\mu^\pm) = \mathcal{B}(K^\pm\to\mu^\pm N)\times\mathcal{B}(N\to\pi^\mp\mu^\pm) \sim
\left|U_{\mu4}\right|^4,
\end{equation*}
where $U_{\mu4}$ is the corresponding mixing parameter. The most stringent published limit on this process is $\mathcal{B}(K^\pm\to\pi^\mp\mu^\pm\mu^\pm) < 1.1\times10^{-9}$ at 90\%~CL~\cite{ba11}. The neutrino decaying as $N\to\pi^\pm\mu^\mp$ can also be observed in the lepton number conserving (LNC) channel $K^\pm\to\pi^\pm\mu^+\mu^-$. The inflaton (or another scalar resonance) in the mass range \(2m_\mu < m_\chi < m_K-m_\pi\) can be produced in a similar way in the decay $K^\pm\to\pi^\pm \chi$ and subsequently decay into a muon pair: $\chi\to\mu^+\mu^-$.

\begin{figure}[t]
\begin{center}
\resizebox{0.5\textwidth}{!}{\includegraphics{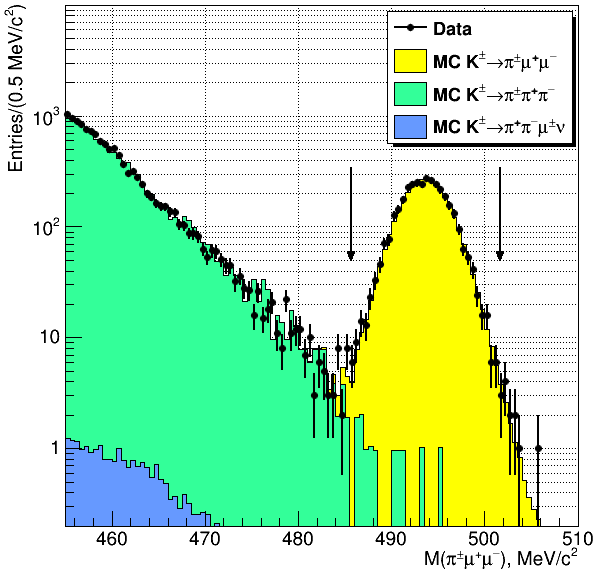}}%
\resizebox{0.5\textwidth}{!}{\includegraphics{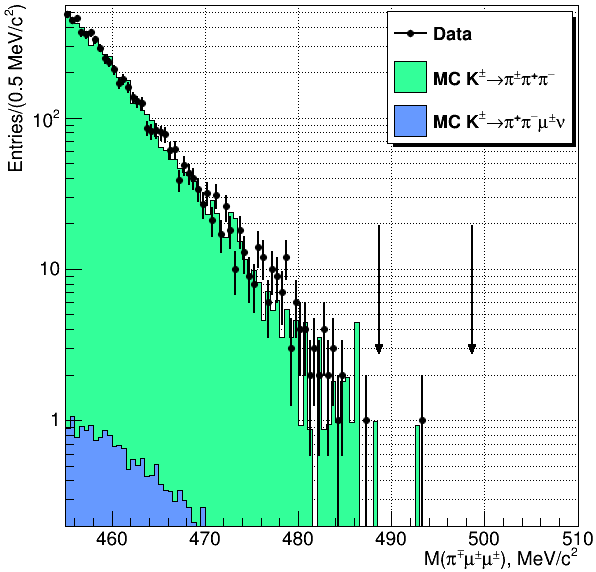}}%
\end{center}
\vspace{-5mm}
\caption{Reconstructed $\pi\mu\mu$ invariant masses for LFC and LFV decays $K^\pm\to\pi^\pm\mu^+\mu^-$ and $K^\pm\to\pi^\mp\mu^\pm\mu^\pm$. Data and simulated events are both shown. The signal mass regions are indicated with vertical arrows.}
\label{fig:pimm}
\end{figure}

The rates of $K^\pm\to\pi^\mp\mu^\pm\mu^\pm$ and $K^\pm\to\pi^\mp\mu^+\mu^-$ decays (collectively referred to as $K_{\pi\mu\mu}$) are measured with respect to the abundant $K^\pm\to\pi^\pm\pi^+\pi^-$ ($K_{3\pi}$) decays. The $K_{3\pi}$ decay also represents the main background via pion decays in flight and mis-identification. The selection of both $K_{\pi\mu\mu}$ and $K_{3\pi}$ decays is based on a three-track vertex topology, with the vertex reconstructed inside the fiducial decay volume. The total reconstructed momentum of the three tracks is required be compatible with the nominal beam momentum, with no transverse momentum with respect to the nominal beam direction. Particle identification is performed using the ratio $E/p$ of the energy deposit measured in the LKr calorimeter to the momentum reconstructed in the spectrometer, as well as the response of the downstream muon detector. The mass spectra of the selected LNC and LFV candidates and the background spectra evaluated with simulations are shown in Fig.~\ref{fig:pimm}.

In the LNV channel, a blind analysis is performed, with the selection optimized and the background simulation validated in the control mass region defined as $M_{\pi\mu\mu}<480~{\rm MeV}/c^2$, where $M_{\pi\mu\mu}$ is the reconstructed invariant mass of the pion and the two muons. The signal region is defined as $|M_{\pi\mu\mu}-m_K|<5~{\rm MeV}/c^2$. For the LNC, the signal region is defined as $|M_{\pi\mu\mu}-m_K|<8~{\rm MeV}/c^2$, and 3489~candidates of the well established flavour changing neutral current decay are observed with sub-percent background contamination. One $K^\pm\to\pi^\mp\mu^\pm\mu^\pm$ candidate event is observed in the signal region while the expected number of background events is \(N_\text{exp} = 1.163\pm0.867_\text{stat}\pm0.021_\text{ext}\pm0.116_\text{syst}\). The upper limit on the branching fraction of the LFV decay is then established to be $\mathcal{B}(K^\pm\to\pi^\mp\mu^\pm\mu^\pm) < 8.6\times 10^{-11}$ at 90\% CL. This limit represents an improvement by more than an order of magnitude with respect to the previous search~\cite{ba11}.

A scan for a resonance of mass \(M_\text{res}\) in the reconstructed \(\pi^\pm\mu^\mp\) invariant mass \(M_{\pi\mu}\) of both LNV and LNC channels, and in the $\mu^+\mu^-$ invariant mass \(M_{\mu\mu}\) of the LNC channel is performed. The step and width of the search windows are determined by the mass resolution \(\sigma({M_\text{res}})\) at the tested value. The mass step between two mass hypotheses is \(0.5\sigma(M_\text{res})\) and the window centered on \(M_\text{res}\) has a width of \(2\sigma(M_\text{res})\). The upper limit on the product of branching fractions at 90\% CL is established for each resonance mass and lifetime hypothesis. As a result of the three-track vertex selection requirement, the acceptances (evaluated with simulations) are inversely proportional to the assumed resonance lifetime for sufficiently large lifetimes.

\begin{figure}[t]
\begin{center}
\resizebox{0.33\textwidth}{!}{\includegraphics{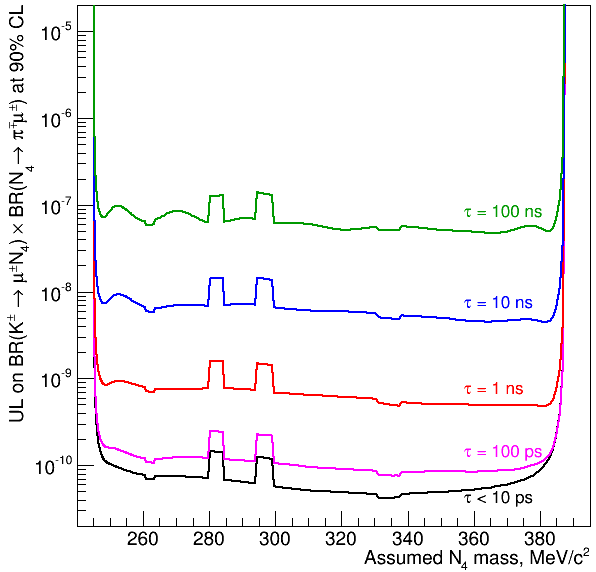}}%
\resizebox{0.33\textwidth}{!}{\includegraphics{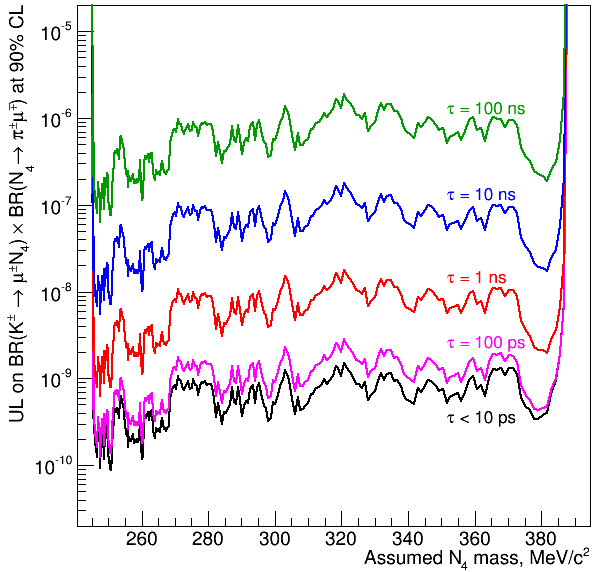}}%
\resizebox{0.33\textwidth}{!}{\includegraphics{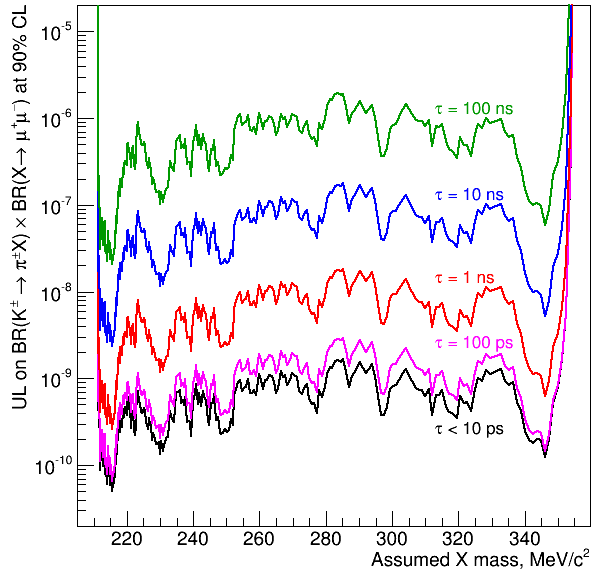}}%
\end{center}
\vspace{-5mm}
\caption{Obtained upper limits at 90\% CL on the products of branching fractions depending on the assumed resonance masses and lifetimes. Left: \(\mathcal{B}(K^\pm\to\mu^\pm N_4)\times\mathcal{B}(N_4\to\pi^\mp\mu^\pm)\); centre: \(\mathcal{B}(K^\pm\to\mu^\pm N_4)\times\mathcal{B}(N_4\to\pi^\pm\mu^\mp)\); right: \(\mathcal{B}(K^\pm\to\pi^\pm N_4)\times\mathcal{B}(N_4\to\mu^\pm\mu^\mp)\).}
\label{fig:lfv_result}
\end{figure}

A total of 284 mass hypotheses are tested on the LNV \(M_{\pi\mu}\) spectrum to search for \(K^\pm\to\mu^\pm N_4;\; N_4\to\pi^\mp\mu^\pm\) decays. In this case two choices are possible to build \(M_{\pi\mu}\), selecting one of the muons or the other. The possibility closest to \(M_\text{res}\) is chosen. No signals are observed, and upper limits of $\mathcal{O}(10^{-10})$ are set on the product of the branching ratios for heavy Majorana neutrinos with lifetimes $\tau_{N_4}<100~{\rm ps}$. The LNC \(M_{\pi\mu}\) spectrum is tested with 280 hypotheses, and no \(K^\pm\to\mu^\pm N_4;\; N_4\to\pi^\pm\mu^\mp\) signals are observed. The obtained upper limits are $\mathcal{O}(10^{-9})$ for sterile neutrinos with lifetimes $\tau_{N_4} < 100~{\rm ps}$. Finally, a scan is performed on the \(M_{\mu\mu}\) spectrum of the LNC channel to search for \(K^\pm\to\pi^\pm \chi;\; \chi\to\mu^\pm\mu^\mp\) with 267 hypotheses, and no signals are observed. The upper limits are $\mathcal{O}(10^{-9})$ for resonance lifetimes $\tau_\chi < 100~{\rm ps}$. The obtained ULs depending on the assumed resonance masses and lifetimes are shown in Fig.~\ref{fig:lfv_result}.

\boldmath
\section{$\pi^0$ electromagnetic transition form factor slope}
\unboldmath
\label{sec:pi0dalitz}
As a $\pi^0$ is produced in four of the six main $K^\pm$ decays, the NA62-$R_K$ experiment exposed to about $2\times10^{10}$ kaon decays in flight in the vacuum fiducial decay region with minimum bias trigger conditions is an ideal environment to study the neutral pion physics. The Dalitz decay $\pi^0_D\to\gamma e^+ e^-$ proceeds through the $\pi^0\gamma\gamma$ vertex with an off-shell photon. The commonly used kinematic variables defined in terms of the
particle four-momenta are:
\begin{equation*}
    x = \left( \frac{M_{e e}}{m_{\pi^0}} \right)^2
    = \frac{(p_{e^+} + p_{e^-})^2}{ m_{\pi^0}^2}, \qquad
    y = \frac{2 \, p_{\pi^0} \cdot \left( p_{e^+} - p_{e^-} \right)}{m_{\pi^0}^2
    (1-x)},
\end{equation*}
where $p_{\pi^0}$, $p_{e^+}$, $p_{e^-}$ are respectively the $\pi^0$ and $e^\pm$ four-momenta, $m_{\pi^0}$ is the mass of the $\pi^0$, and $M_{ee}$ is the $e^+e^-$ invariant mass. The physical region is given by
\begin{equation*}
    r^2 = \left(\frac{2 m_e}{m_{\pi^0}}\right)^2 \leq x \leq 1, \quad |y| \leq
    \sqrt{1 - \frac{r^2}{x}} \;.
\end{equation*}
The $\pi^0_D$ differential decay width normalised to the $\pi^0_{2\gamma}\to\gamma\gamma$ decay width reads
\begin{equation*}
    \frac{1}{\Gamma(\pi^0_{2\gamma})} \frac{\text{d}^2 \Gamma(\pi^0_D)}{\text{d}x \text{d}y} =
    \frac{\alpha}{4\pi} \frac{(1-x)^3}{x} \left(1 + y^2 + \frac{r^2}{x}\right) \; \left(1+\delta(x,y)\right) \;
    \left|\mathcal{F}(x)\right|^2,
\end{equation*}
where $\mathcal{F}(x)$ is the semi-off-shell electromagnetic transition form factor (TFF) of the $\pi^0$ and $\delta(x,y)$ encodes the radiative corrections.

The TFF is usually expanded as $\mathcal{F}(x) = 1+ax$ where $a$ is the form factor slope parameter. This approximation is justified by the smallness of this parameter. In the vector meson dominance (VMD) model, it is dominated by the $\rho$ and $\omega$ mesons, resulting in a slope of $a\approx m^2_{\pi^0} \left(m_\rho^{-2}+m_\omega^{-2}\right)/2 \approx 0.03$. 

A crucial aspect of measuring the $\pi^0$ TFF with the $\pi^0_D$ decays is control of the radiative corrections in the differential rate: their effect on the differential decay rate exceed that of the TFF. The first study of the corrections on the differential rate was performed in~\cite{Lautrup1971} in the soft-photon approximation, and extended in~\cite{Mikaelian1972}. A recent improved computation~\cite{Husek2015_rad} largely triggered by and used for this analysis includes additional second-order contributions and accounts for the emission of radiative photons, i.e. the internal bremsstrahlung contribution.

\begin{figure}[h]
\begin{center}
\resizebox{0.47\textwidth}{!}{\includegraphics{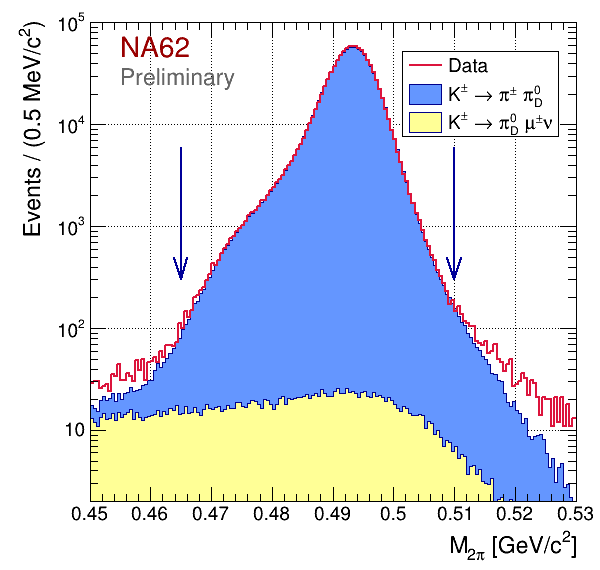}}%
\resizebox{0.53\textwidth}{!}{\includegraphics{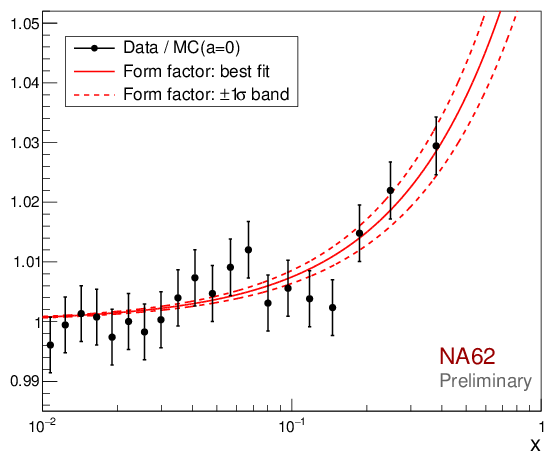}}
\end{center}
\vspace{-9mm}
\caption{Left: reconstructed $\pi^\pm\pi^0_D$ mass spectrum of the data and simulated samples. The residual difference is mainly due to the non-inclusion of the $K^\pm\to\pi^0_D e^\pm\nu$ component of the signal into the simulation. Right: Illustration of $\pi^0$ the TFF fit result showing the data/MC ratio, with the MC sample weighted to obtain $a=0$. The events are divided into 20 equipopulous bins, and the markers are located at the bin barycentres of each bin. The solid line represents the TFF function with a slope value equal to the fit central value. The dashed lines indicate the $1\sigma$ band.}
\vspace{-5mm}
\label{fig:pi0d_result}
\end{figure}

A sample of $1.05\times 10^6$ $K^\pm\to\pi^\pm\pi^0_D$ decays with negligible background has been selected in the NA62-$R_K$ data using a spectrometer three-track vertex selection and reconstructing the photon in the LKr calorimeter. A $\chi^2$ fit to the reconstructed $x$ distribution of the data $\pi^0_D$ candidates with equipopulous binning has been performed to extract the TFF slope value (Fig.~\ref{fig:pi0d_result}). The main systematic uncertainties arise from the simulation of the beam spectrum and from the calibration of the spectrometer global momentum scale. The preliminary result is
\begin{equation*}
a = (3.70 \pm 0.53_\text{stat} \pm 0.36_\text{syst})\times 10^{-2},
\end{equation*}
with $\chi^2/\text{ndf} = 52.5/49$ corresponding to a \textit{p}-value of 0.34. This measurement represents an observation of a positive $\pi^0$ electromagnetic TFF slope (with more than $5\sigma$ significance) in the time-like region of momentum transfer.

It is interesting to note that the $\pi^0_D$ spectrum has been used to search for ``dark sector'' Standard Model extension. In particular, a search for the dark photon (DP) production in the $\pi^0\to\gamma A'$ decay followed by the prompt $A'\to e^+e^-$ decay has been performed with the $\pi^0_D$ data sample collected by the NA48/2 experiment~\cite{ba15}. No DP signal has been observed, providing improved upper limits on the mixing parameter $\varepsilon^2$ in the mass range 9--70 MeV/$c^2$. In combination with other experimental searches, this result ruled out the DP as an explanation for the muon $(g-2)$ measurement under the assumption that the DP couples to quarks and decays predominantly to SM fermions.

\section*{Conclusions}

Searches for the LNV $K^\pm\to\pi^\mp\mu^\pm\mu^\pm$ decay and resonances in \(K^\pm\to\pi\mu\mu\) decays using the 2003--04 data sample collected by the NA48/2 experiment at CERN are presented. A new upper limit at 90\% CL on the branching ratio of the LNV channel is established: ${\cal B}<8.6\times 10^{-11}$, improving the previous best limit by more than an order of magnitude. Upper limits are set on the products of branching ratios $\mathcal{B}(K^\pm\to\mu^\pm
N_4)\mathcal{B}(N_4\to\pi^\mp\mu^\pm)$, $\mathcal{B}(K^\pm\to\mu^\pm
N_4)\mathcal{B}(N_4\to\pi^\pm\mu^\mp)$ and $\mathcal{B}(K^\pm\to\pi^\pm\chi)\mathcal{B}(\chi\to\mu^+\mu^-)$ as
functions of the assumed resonance masses and lifetimes. These limits are in the $10^{-9}$ to $10^{-10}$ range for resonance lifetimes below 100~ps.

A preliminary result from a measurement of the electromagnetic transition form factor slope of the \(\pi^0\) with $1.05\times 10^6$ \(\pi^0_D\to e^+e^-\gamma\) reconstructed decays collected at the NA62 experiment at CERN in 2007 is also reported. The obtained result $a = (3.70\pm 0.64)\times 10^{-2}$ is the first observation of positive TFF slope in the time-like transfer momentum region.


\end{document}